\documentclass[reprint,superscriptaddress,amsmath,amssymb,aps,prb]{revtex4-2}

\usepackage{graphicx}
\usepackage{dcolumn}
\usepackage{bm}
\usepackage{braket}
\usepackage{multirow}

\usepackage[english]{babel}
\newcommand*\diff{\mathop{}\!\mathrm{d}}

\usepackage[table]{xcolor}
\usepackage{hyperref}
\hypersetup{
	colorlinks   = true, 
	urlcolor     = blue,
	linkcolor    = blue, 
	citecolor   = blue 
}

\begin{document}

\preprint{APS/123-QED}

\title{Accurate and efficient structure factors in \\ ultrasoft pseudopotential and projector augmented wave DFT}

\author{Benjamin X. Shi}
\email{mail@benjaminshi.com}
\author{Rebecca J. Nicholls}
\author{Jonathan R. Yates}%
\affiliation{Department of Materials, University of Oxford, Parks Road, Oxford, OX1 3PH, United Kingdom}

\date{\today}

\begin{abstract}
Structure factors obtained from diffraction experiments are one of the most important quantities for characterizing the electronic and structural properties of materials.
Methods for calculating this quantity from plane-wave density functional theory (DFT) codes are typically prohibitively expensive to perform, requiring the electron density to be constructed and evaluated on dense real-space grids. 
Making use of the projector functions found in both the Vanderbilt ultrasoft pseudopotential and projector augmented wave methods, we implement an approach to calculate structure factors which avoids the use of a dense grid by separating the rapidly changing contributions to the electron density and treating them on logarithmic radial grids. 
This approach is successfully validated against structure factors obtained from all-electron DFT and experiments for three prototype systems, allowing structure factors to be obtained at all-electron accuracy at a fraction of the cost of previous approaches for plane-wave DFT.
\end{abstract}

\maketitle


\section{\label{sec:intro} Introduction}

The structure factor (SF) is a fundamentally important quantity in the physical and biological sciences.
Experimental SFs have become the principal tool for characterizing the geometrical structure (e.g.,\ atomic positions and elements) of a range of crystalline matter, from simple materials~\cite{groomCambridgeStructuralDatabase2016} to complex bio-molecules~\cite{bermanProteinDataBank2000}. 
Beyond structural information, the (x-ray) SFs are the Fourier coefficients of the electron density (ED) - $n(\mathbf{r})$ - and allow for its reconstruction through an inverse Fourier transform~\cite{hammondBasicsCrystallographyDiffraction2015a}. 
The ED contains a wealth of information, as established by the Hohenberg-Kohn theorems~\cite{hohenbergInhomogeneousElectronGas1964}, which states that the ground-state electronic properties of a system are a unique functional of the ground-state ED. 
Reconstructed EDs have been used to investigate the properties of a range of materials, from the anisotropic elastic constants of Al~\cite{nakashimaBondingElectronDensity2011a} to the electronic origins of high-temperature cuprate superconductors~\cite{zuoDirectObservationDorbital1999,forganMicroscopicStructureCharge2015}.

\textit{Accurate} SFs calculated from first-principles methods, particularly density functional theory (DFT) are necessary for the above applications of SFs.
For example, SFs computed from DFT have been used to augment experimental SFs to allow full reconstruction of the ED. By themselves, SFs obtained from x-ray~\cite{moriElectronDensityDistribution2002,forganMicroscopicStructureCharge2015,aubertPeriodicProjectorAugmented2011b}, $\gamma$-ray~\cite{jauchElectronDensityCubic2005,jauchElectronDensityDistribution2009,jauchElectronDensityDistribution2011} or electron diffraction~\cite{shibataMolecularElectronDensity1999,zuoChargeDensityMgO1997b,ogataDeterminationElectrostaticPotential2008,pengMeasuringDensityFunctional2021} experiments can only provide a finite set of SFs, introducing Fourier series truncation errors to the reconstructed ED~\cite{wuValenceelectronDistributionMgB2004} if used alone.
Many of these diffraction experiments are also incapable of obtaining the phase of the complex SFs in non-centrosymmetric crystals -- the phase problem~\cite{taylorPhaseProblem2003a}. 
DFT helps to alleviate these problems as it can generate many SFs to augment those that are not provided by experiments while also providing phase information.

Accurate SFs from DFT are also useful for assessing the quality of experimental diffraction techniques, which can suffer from problems afflicting their accuracy.
For example, extinction effects~\cite{howardXrayDiffractionStudy1977a,schmidtElectronDensityStudies1985a} and the source of x-rays~\cite{schmokelComparativeStudyXray2013} can affect measurements in x-ray diffraction experiments.
In electron diffraction, material preparation~\cite{pengHowSpecimenPreparation2017} and instrument distortions~\cite{gruene3DElectronDiffraction2021} can serve as potential error sources.

High-precision SFs from experiments can also help validate the approximations used to make DFT calculations computationally tractable.
Based upon the Hohenberg-Kohn theorems, practical DFT calculations attempt to approximate the exact energy functional, which is unknown, to the ED through density functional approximations (DFAs). 
There is a whole `zoo'~\cite{burschBestPracticeDFTProtocolsa} of available DFAs, with no systematic manner to determine their accuracy. 
In recent years, there has been evidence~\cite{medvedevDensityFunctionalTheory2017c} to show that although modern DFAs give improved energetic descriptions of (atomic) systems, the description of the ED is worse. 
This deficiency arises because many modern DFAs have been constructed through empirical fitting of reference energies, typically neglecting the ED due to the lack of available references.  
Towards this end, high-precision SFs obtained from diffraction experiments can be compared to those obtained from various DFAs to assess their relative accuracy, and it has been successfully performed for several materials~\cite{zuoTheoreticalChargeDensity1997a,friisMagnesiumComparisonDensity2003b,zuoChargeDensityMgO1997b,saeterliExperimentalTheoreticalStudy2011}.

Beyond the need for accurate SFs from DFT, it is also highly desirable that they can be obtained \textit{efficiently} without incurring heavy computational burden/time to allow for more complex systems to be tackled. 
Unfortunately, current approaches to calculate the SF from DFT methods are not efficient, requiring a high computational cost. 
There are an array of DFT methods, but the key methods used for computing the SF are either all-electron (AE) or plane-wave pseudopotential (PP) DFT methods. 
AE DFT approaches treat all the electrons in the system explicitly, enabling highly accurate calculations. 
This approach is the most common approach for computing SFs due to the ease at obtaining the SF from the ED.
However, it comes at significant computational cost due to the $\sim O(N^3)$~\cite{aaronsPerspectiveMethodsLargescale2016} scaling of DFT. 
%

In PP DFT, the inclusion of a pseudopotential~\cite{schwerdtfegerPseudopotentialApproximationElectronic2011a} means that only valence electrons need to be treated, decreasing the computational cost significantly. 
The resulting ED from the PP DFT SCF calculation cannot be directly used to obtain the SF (as in AE DFT) as it includes only valence electrons and has been smoothed -- pseudized -- near the core.
In modern plane-wave DFT codes, the Vanderbilt ultrasoft pseudopotential~\cite{vanderbiltSoftSelfconsistentPseudopotentials1990d} or the projector augmented wave (PAW)~\cite{blochlProjectorAugmentedwaveMethod1994c,kresseUltrasoftPseudopotentialsProjector1999c} methods can be employed to restore the AE total ED from the pseudized (PS) valence ED by adding a compensating augmentation charge and including the (frozen-)core ED. 
In the typical plane-wave DFT codes used, the AE total ED must be reconstructed on a real-space fast Fourier transform (FFT) grid. 
This FFT grid has to be several orders of magnitude denser than the default used for the PS valence ED to have sufficient spatial resolution to accurately capture the rapid oscillations of the AE total ED near the nucleus. 
As a result, the calculations require a large amount of memory and time, with several studies explicitly highlighting the difficulty with converging the total ED to a sufficient precision due to computational limitations~\cite{saeterliExperimentalTheoreticalStudy2011,flage-larsenBondAnalysisPhosphorus2010}.

In this paper, we propose a highly efficient and accurate method of calculating the AE SF from PP DFT that can be implemented in both the Vanderbilt ultrasoft pseudopotential and PAW methods. 
It works by removing the rapidly changing contributions (i.e. core and augmentation charges) from the AE total ED and treating their contributions to the AE SF separately.
With this change, the FFT grid does not need to be increased beyond its default size since it only represents the PS valence ED. 
The core and augmentation contributions to the ED, which exist only within a small region around the atom, will have their radial components treated on atom-centered logarithmic radial support grids, while the angular components are treated analytically. 
As the radial support grids are one-dimensional, they can be made very dense near the origin where the ED contributions vary the most, allowing for high accuracy at a low computational cost. 
This method is successfully validated against AE DFT and experiment for a range of materials.


\section{\label{sec:aedensity} The AE Electron Density}

This section details how the AE ED is obtained from PP DFT calculations using the Vanderbilt ultrasoft pseudopotential and PAW methods. 
The theory is applicable for both methods, so unless otherwise stated, PAW DFT will be used to denote both methods hereafter.

In PAW DFT, the AE wave function $\psi_n (\mathbf{r})$ for each of the $n$ valence (Kohn-Sham) orbitals can be reconstructed from its corresponding pseudized (PS) wave function $\tilde{\psi}_n (\mathbf{r})$ through a linear transformation~\cite{blochlProjectorAugmentedwaveMethod1994c}:
\begin{equation} \label{eq:valenceed}
\psi_n(\mathbf{r}) = \tilde{\psi}_n(\mathbf{r}) + \sum_{\mathbf{R}ju}  \left [ \phi^j_{u} (\mathbf{r}_\mathbf{R}^j) - \tilde{\phi}^j_u (\mathbf{r}_\mathbf{R}^j) \right ] \braket{\tilde{p}^j_{u} | \tilde{\psi}_n}.
\end{equation}
Within this expression, $\phi^j_{u} (\mathbf{r}_\mathbf{R}^j)$ and $\tilde{\phi}^j_u (\mathbf{r}_\mathbf{R}^j)$  are the AE and PS partial waves respectively for each atom $j$ in the unit cell, with the projectors $\bra{\tilde{p}^j_{u}}$ designed to be dual to the PS partial waves: $\braket{\tilde{p}^{j}_{u} | \tilde{\phi}^{j'}_{u'}} = \delta_{jj'} \delta_{uu'}$. 
The AE partial waves are a set of wave functions obtained from the corresponding reference atom~\cite{koellingTechniqueRelativisticSpinpolarised1977a}, where $u$ is the composite index for the angular momentum quantum numbers $l,m$ as well as an index $k$ to label partial waves constructed at different reference energies~\cite{blochlProjectorAugmentedwaveMethod1994c}. The vectors
\begin{equation} 
\mathbf{r}_\mathbf{R}^j = \mathbf{r} - \mathbf{r}_j - \mathbf{R}
\end{equation}
are used to denote the spatial dependence of the partial waves and projectors to emphasize that these functions are atom-centered. 
These equations assume a periodic material, where $\mathbf{R}$ are the (infinite) set of lattice vectors and $\mathbf{r}_j$ denotes the position of atom $j$ in the unit cell.
Computationally, the partial waves and projectors can be expressed as a radial function (stored on logarithmic radial grids) multiplied by a spherical harmonic:
\begin{equation} \label{eq:partialwaves}
\phi_u (\mathbf{r}) = R_{lk} (r) Y_{l m} (\hat{\mathbf{r}}),
\end{equation}
with the PS partial waves differing from the AE partial waves only within a cutoff $r_c^l$, where its radial component has been pseudized.

The AE valence ED $n_\text{val}(\mathbf{r})$ can be given as a sum of two contributions:
\begin{equation}
n_\textrm{val}(\mathbf{r}) = \tilde{n}_\textrm{val} (\mathbf{r}) + n_\textrm{aug}(\mathbf{r}).
\end{equation}
The first term is the PS valence ED, which is the ED resulting from the PS (valence) wave functions in Eq.~\ref{eq:valenceed}:
\begin{equation}
\tilde{n}_\textrm{val}(\mathbf{r}) = \sum_n f_n |\tilde{\psi}_n (\mathbf{r})|^2,
\end{equation}
where $f_n$ are the occupation numbers. 
The FFT grid is designed to store this smooth function, with its default size sufficient to sample and represent it fully. 
The second term is the augmentation charge, which restores the PS valence ED to the AE valence ED, taking the form:
\begin{equation}
n_\text{aug}(\mathbf{r}) =  \sum_{\mathbf{R}j u_1 u_2} \rho_{u_1 u_2}^j Q_{u_1 u_2}^j (\mathbf{r}_\mathbf{R}^j).
\end{equation}
The augmentation functions $ Q_{u_1 u_2}^j (\mathbf{r}_\mathbf{R}^j)$ are defined as
\begin{equation} \label{eq:augmentationfunction}
Q_{u_1 u_2}^j (\mathbf{r}) = \phi^j_{u_1}(\mathbf{r})^* \phi^{j}_{u_2}(\mathbf{r})  - \tilde{\phi}^j_{u_1}(\mathbf{r})^* \tilde{\phi}^{j}_{u_2}(\mathbf{r}),
\end{equation}
where $u_1$ and $u_2$ are two sets of $u$ indices, with $\rho_{u_1 u_2}^{j}$ giving the occupancy of each $u_1,u_2$ augmentation function channel for atom $j$:
\begin{equation}
\rho_{u_1 u_2}^{j} = \sum_n f_n \braket{\tilde{\psi}_n | \tilde{p}_{u_1}^j} \braket{ \tilde{p}_{u_2}^j | \tilde{\psi}_n}.
\end{equation}
The augmentation function $Q_{u_1 u_2}^j (\mathbf{r})$ will be localized around atom $j$ in an `augmentation' region. The two most common types of pseudopotentials are the norm-conserving PPs (NCPs) and ultrasoft PPs (USPs). 
NCPs are constructed to preserve the norm of the wave (i.e. the integral of the augmentation functions within the augmentation region is zero by construction), while USPs relax this condition, requiring fewer plane-waves to describe the PS valence wave functions at the cost of additional complexity. 
It is standard practice when using NCPs to neglect any augmentation to the charge density during the calculation of the ground-state.
Following from Eqs.~\ref{eq:partialwaves} and \ref{eq:augmentationfunction}, $Q_{u_1 u_2}^j (\mathbf{r})$ can be expressed as the product of a radial function $\Delta R_{u_1 u_2}^j (r)$ and two spherical harmonics:
\begin{equation} \label{eq:augmentationfunction1}
Q_{u_1 u_2}^j (\mathbf{r}) = \Delta R_{u_1 u_2}^j (r) Y^*_{l_1 m_1} (\hat{\mathbf{r}}) Y_{l_2 m_2} (\hat{\mathbf{r}}).
\end{equation}

Within the PAW method, the radial functions in the augmentation functions are treated on atom-centered logarithmic radial support grids~\cite{kresseUltrasoftPseudopotentialsProjector1999c}, with the spherical harmonics treated analytically, while in the Vanderbilt ultrasoft pseudopotential approach, these functions are typically pseudized and placed onto the FFT grid~\cite{laasonenCarParrinelloMolecularDynamics1993}. 
While this pseudization gives accurate total energies~\cite{lejaeghereReproducibilityDensityFunctional2016a}, the use of a PS compared to the AE augmentation charge introduces significant errors in the calculations of SF, as will be shown in Sec.~\ref{sec:validation}.

The PAW method typically utilizes the frozen core approximation. Under this scheme, the core ED is a superposition of the core EDs $\rho_\textrm{core}^j$ obtained from isolated atoms:
\begin{equation} \label{eq:coreed}
n_\textrm{core}(\mathbf{r}) = \sum_{\mathbf{R}j} \rho^j_\textrm{core} (|\mathbf{r}_\mathbf{R}^j|).
\end{equation}
These atomic core EDs are spherically symmetric and obtained from solving the radial Kohn-Sham Schr\"{o}dinger equation~\cite{koellingTechniqueRelativisticSpinpolarised1977a} on the logarithmic radial support grids. 

Overall, the total AE ED will then be the sum of the PS valence, augmentation charge and frozen-core EDs:
\begin{equation} \label{eq:totaled}
n(\mathbf{r}) = \tilde{n}_\textrm{val}(\mathbf{r}) + n_\textrm{aug}(\mathbf{r}) + n_\textrm{core}(\mathbf{r}).
\end{equation}

\section{\label{sec:aesf} The AE Structure Factors}

This section details how the AE SFs can be obtained from the AE ED in PAW DFT codes. 
The AE SFs $F(\mathbf{H})$ are the Fourier coefficients of the AE (total) ED:
\begin{equation}
F(\mathbf{H}) = \mathcal{F}[n(\mathbf{r})],
\end{equation}
where $\mathbf{H} = h \mathbf{a}^* + k \mathbf{b}^* + l\mathbf{c}^*$ is the scattering vector corresponding to the $(hkl)$ plane and the crystallographic convention for the primitive reciprocal lattice vectors $\mathbf{a}^*, \mathbf{b}^*$ and $\mathbf{c}^*$ has been used. 
Current methods of obtaining the AE SFs from the AE ED in PAW DFT codes involves constructing the AE ED onto a uniform grid before applying a fast Fourier transform (FFT).
This approach is highly inefficient because the FFT grid used has to be several orders of magnitude denser than the default (designed for the PS valence ED only) to accommodate the rapidly varying augmentation and core charges near the nuclei.
As shown in Fig.\ S1 of the Supplemental Material~\cite{sm}, this leads to orders of magnitude increases in time and peak RAM. 
The impact of the resulting additional computational burden has been noted in previous studies~\cite{saeterliExperimentalTheoreticalStudy2011}.

In this section, we propose and derive a new approach to calculate AE SFs efficiently and accurately in the PAW DFT method. 
It works by separating the three individual contributions to the AE total ED from PAW DFT and treating the two terms that require high spatial resolution near the nuclei: $n_\textrm{aug}$ and $n_\textrm{core}$, on logarithmic radial support grids and analytically for radial and angular components respectively. 
These radial grids are one-dimensional with a high density of points placed near the nuclei to achieve high accuracy efficiently.

From the linearity principle of the Fourier transform, the SF can be separated into three contributions:
\begin{equation} \label{eq:totalsf}
F(\mathbf{H}) = \tilde{F}_\textrm{val}(\mathbf{H}) + F_\textrm{core}(\mathbf{H}) + F_\textrm{aug}(\mathbf{H}),
\end{equation}
where $\tilde{F}_\textrm{val}(\mathbf{H}), F_\textrm{core}(\mathbf{H})$ and $F_\textrm{aug}(\mathbf{H})$ are the Fourier transforms of $\tilde{n}_\textrm{val}(\mathbf{r}), n_\textrm{core}(\mathbf{r})$ and $n_\textrm{aug}(\mathbf{r})$ respectively. 
By nature, $\tilde{n}_\textrm{val}(\mathbf{r})$ is constructed to be fully described on the coarse (default) FFT grid size. 
Thus, its Fourier transform can be computed efficiently using the FFT method, a fundamental component of all plane-wave DFT codes.

The theory for treating the core and augmentation contributions to the SF in the next two subsections relies on some of the methods developed for the independent atom model (IAM) -- commonly used in crystallography.
The IAM is constructed as a summation of isolated atomic densities $\rho^j (\mathbf{r})$ about their atomic positions:
\begin{equation} \label{eq:iam}
n_\textrm{IAM}(\mathbf{r}) = \sum_{\mathbf{R}j} \rho^j (\mathbf{r}_\mathbf{R}^j),
\end{equation}
and it is possible to show that its SF takes the form~\cite{hammondBasicsCrystallographyDiffraction2015a}:
\begin{equation}
F_\textrm{IAM}(\mathbf{H}) = \sum_{j} f^{j} (\mathbf{H}) \exp(i 2\pi \mathbf{H} \cdot \mathbf{r}_{j}),
\end{equation} 
where $f^j(\mathbf{H})$ is the atomic scattering factor, defined to be the Fourier transform of the corresponding $\rho^j(\mathbf{r})$:
\begin{equation} \label{eq:iam3dformfactor}
f^j (\mathbf{H}) = \int \rho^j (\mathbf{r}) \exp(i 2 \pi \mathbf{H} \cdot \mathbf{r}) \diff{\mathbf{r}}.
\end{equation}

\subsection{Core contribution}

Recalling Eq.~\ref{eq:coreed}, the core ED is a summation of atom-centered densities, much like the IAM. Thus, using the SF expressions derived for the IAM, the SF contribution from the core ED takes the form:
\begin{equation}
F_\textrm{core}(\mathbf{H}) = \sum_{j} f^{j}_\textrm{core} (\mathbf{H}) \exp(i 2\pi \mathbf{H} \cdot \mathbf{r}_{j}),
\end{equation} 
where the atomic core scattering factor $f^{j}_\textrm{core}(\mathbf{H})$ is the Fourier transform of the corresponding $\rho^j_\textrm{core}(\mathbf{r})$:
\begin{equation} \label{eq:core3dformfactor}
f_\textrm{core}^j (\mathbf{H}) = \int \rho^j_\textrm{core} (|\mathbf{r}|) \exp(i 2 \pi \mathbf{H} \cdot \mathbf{r}) \diff{\mathbf{r}}.
\end{equation}

As $\rho_\textrm{core}(|\mathbf{r}|)$ is a spherically symmetric function about the origin, it is more convenient to use the spherical polar coordinate system. 
Within this coordinate system, the plane-wave function can be expanded into complex spherical harmonic functions~\cite{shmueliInternationalTablesCrystallography2001}:
\begin{equation} \label{eq:planewaveexpansion}
\exp(i 2 \pi  \mathbf{H} \cdot \mathbf{r}) = 4 \pi \sum_{l = 0}^\infty \sum_{m = -l}^l i^{l} j_l(2 \pi Hr) Y_{l,m} (\hat{\mathbf H}) Y_{l,m}^{*}(\hat{\mathbf r}),
\end{equation}
where $j_l (r)$ is the spherical Bessel function of order $l$. Substituting Eq.~\ref{eq:planewaveexpansion} into Eq.~\ref{eq:core3dformfactor} would then yield:
\begin{equation} \label{eq:coreformfactor}
f_\textrm{core}^j (H) = \int \rho_\textrm{core}^j (r)  4 \pi r^2  j_0(2 \pi H r) \diff r,
\end{equation}
as only the $l=0$ terms persist due to the rotational invariance of $\rho_\textrm{core}^j (r)$ (and in turn $f_\textrm{core}^j (H)$).
This equation can be evaluated on the native logarithmic radial grids for each atom $j$ within the unit cell and as a result, the core contribution to the SF no longer needs to be evaluated on the FFT grid.

\subsection{Augmentation contribution}

Like the core ED and IAM, the augmentation charge also consists of a summation of atom-centered functions, so it can be written to take the form:
\begin{equation}
F_{\textrm{aug}}(\mathbf{H}) = \sum_j \exp(i 2 \pi \mathbf{H} \cdot \mathbf{r}_j) \sum_{u_1 u_2} f^j_{u_1 u_2} (\mathbf{H}),
\end{equation}
where
\begin{equation} \label{eq:augmentationformfactor}
f^j_{u_1u_2} (\mathbf{H}) = \rho_{u_1 u_2}^j \int Q_{u_1 u_2}^j(\mathbf{r}) \exp(i 2 \pi \mathbf{H} \cdot \mathbf{r}) \diff \mathbf{r}.
\end{equation}

The augmentation charge differs from the core ED in that these atom-centered functions $Q_{u_1 u2}^j(\mathbf{r})$ are not spherically symmetric about the origin, instead being expressed as a multiple of a radial function and two spherical harmonics (see Eq.~\ref{eq:augmentationfunction1}). 
The spherical harmonics form a complete set, so their products can be expressed as an expansion of single spherical harmonics:
\begin{equation}
Y_{l_1 m_1}^{\ast}  (\hat{\mathbf{r}}) Y_{l_2 m_2} (\hat{\mathbf{r}}) = \sum_L C_{l_1 l_2 L}^{m_1 m_2} Y_{L M}  (\hat{\mathbf{r}}),
\end{equation}
where  $M=-m_1 + m_2$, $ |l_1 - l_2| \leq L \leq l_1 + l_2$ and $C_{l_1 l_2 L}^{m_1 m_2}$ are the Clebsch-Gordan coefficients. 
The augmentation functions $Q_{u_1 u2}^j(\mathbf{r})$ can then be rewritten as
\begin{equation} \label{eq:augmentationfunction2}
Q_{u_1 u2}^j(\mathbf{r}) = \Delta R_{u_1 u_2}^{j}(r) \sum_L C_{l_1 l_2 L}^{m_1 m_2} Y_{L M}  (\hat{\mathbf{r}}).
\end{equation}
When this new formulation of $Q_{u_1 u_2}^j(\mathbf{r})$ is substituted into Eq.~\ref{eq:augmentationformfactor}, $f^j_{u_1u_2} (\mathbf{H})$ can be further simplified to
\begin{equation} \label{eq:augmentationformfactor2}
\begin{split}
f_{u_1 u_2}^j (\mathbf{H})  =  \rho_{u_1 u_2}^j \sum_{L}  C_{l_1 l_2 L}^{m_1 m_2} Y_{L M} (\hat{\mathbf{H}})  4 \pi i^{L} \\
\times \int j_{L} (2 \pi H r) \Delta R_{u_1 u_2}^{j}(r) r^2 \diff r,
\end{split}
\end{equation}
where the expansion of the plane-wave in spherical harmonics (Eq.~\ref{eq:planewaveexpansion}) and subsequently the orthonormal properties of complex spherical harmonics has been used to arrive at the final expression. 
Within Eq.~\ref{eq:augmentationformfactor2}, the $\int j_{L} (2 \pi H r) \Delta R_{u_1 u_2}^{j}(r) r^2 \diff r$ integral can be evaluated to high precision on the logarithmic radial support grids, while the angular components are treated analytically using spherical harmonics, removing the need to evaluate the augmentation contribution on the FFT grid.

With the method outlined in this section, only the default FFT grid size has to be used to evaluate converged AE SFs in PAW DFT since it only houses the PS valence ED, with the other terms being evaluated efficiently on radial support grids, resulting in orders of magnitude savings in computational cost (see Fig.\ S1 of the Supplemental Material~\cite{sm}).

\section{\label{sec:validation} Validation against AE DFT}

To ensure that the approach outlined in Sec.~\ref{sec:aesf} is accurate, we have compared the SFs produced via this approach in PAW DFT against AE DFT calculations for the same DFA for three prototypical systems.

\subsection{Computational details}
The approach was implemented in CASTEP~\cite{clarkFirstPrinciplesMethods2005a}, which is a plane-wave PP DFT code that can use either NCP or USP for ground-state calculation and the PAW method as a post-processing approach to calculate properties such as hyperfine tensors, NMR properties~\cite{yatesCalculationNMRChemical2007b} and now SFs. 
We refer to AE SFs obtained from using this full approach (with augmentation charges treated on radial support grids) as AE-USP or AE-NCP depending on the class of PP used. 
To investigate the importance of using the full valence ED, results will also be presented for what we shall refer to as the PS-USP and PS-NCP approaches. 
The PS-NCP approach uses the pseudized valence ED and the ED from the core electrons, without any augmentation charge. 
The PS-USP also includes the contribution from the pseudized augmentation charge evaluated on the same FFT grid as the valence ED. 
Details of the pseudopotentials used are given in Sec.\ II of the Supplemental Material~\cite{sm}, with all input parameters and output files from this work made available online (see data availability statement at end of paper).

The AE DFT SF calculations are performed using the WIEN2k code~\cite{blahaWIEN2kAPWProgram2020c}, which utilizes a APW+lo basis set.
For both AE and PAW DFT calculations, the energy cutoff ($R_\textrm{MT}K$ for AE DFT) and number of $k$-points were varied to ensure that the SFs were numerically converged to 0.001 (e)lectrons, the limits of experimental accuracy. 
SFs were compared for both the LDA and PBE DFAs in three materials that highlight different bonding and unit cell systems: diamond Si (covalent), hcp Mg (metallic) and rocksalt MgO (ionic). 
In each system, the amplitudes of SFs with scattering length $\mathbf{s} = |\mathbf{H}|/2$ below 1.5 \AA{}\textsuperscript{-1} were analyzed, removing any SFs that have equivalent amplitudes due to symmetry or under Friedel's law~\cite{coppensXrayChargeDensities1997}. 
All forbidden reflections, under the spherical IAM, were also removed, except for the (222) reflection in Si as it has a noticeable amplitude due to the asphericity of the atoms~\cite{braggIntensityXrayReflection1920}. 
The $R$-factor is used to assess the accuracy of our calculations relative to a reference and is defined as:
\begin{equation}
R^\textrm{ref.} = \dfrac{\sum_{\mathbf{H}} |F^\textrm{ref.}(\mathbf{H}) -  F^\textrm{calc.} (\mathbf{H})|}{\sum_{\mathbf{H}} |F^\textrm{ref.} (\mathbf{H})|},
\end{equation}
where the reference will be either WIEN2k calculations in this section or experimental values in the following section.

\subsection{Results and discussion}
\begin{figure}[b]
	\includegraphics{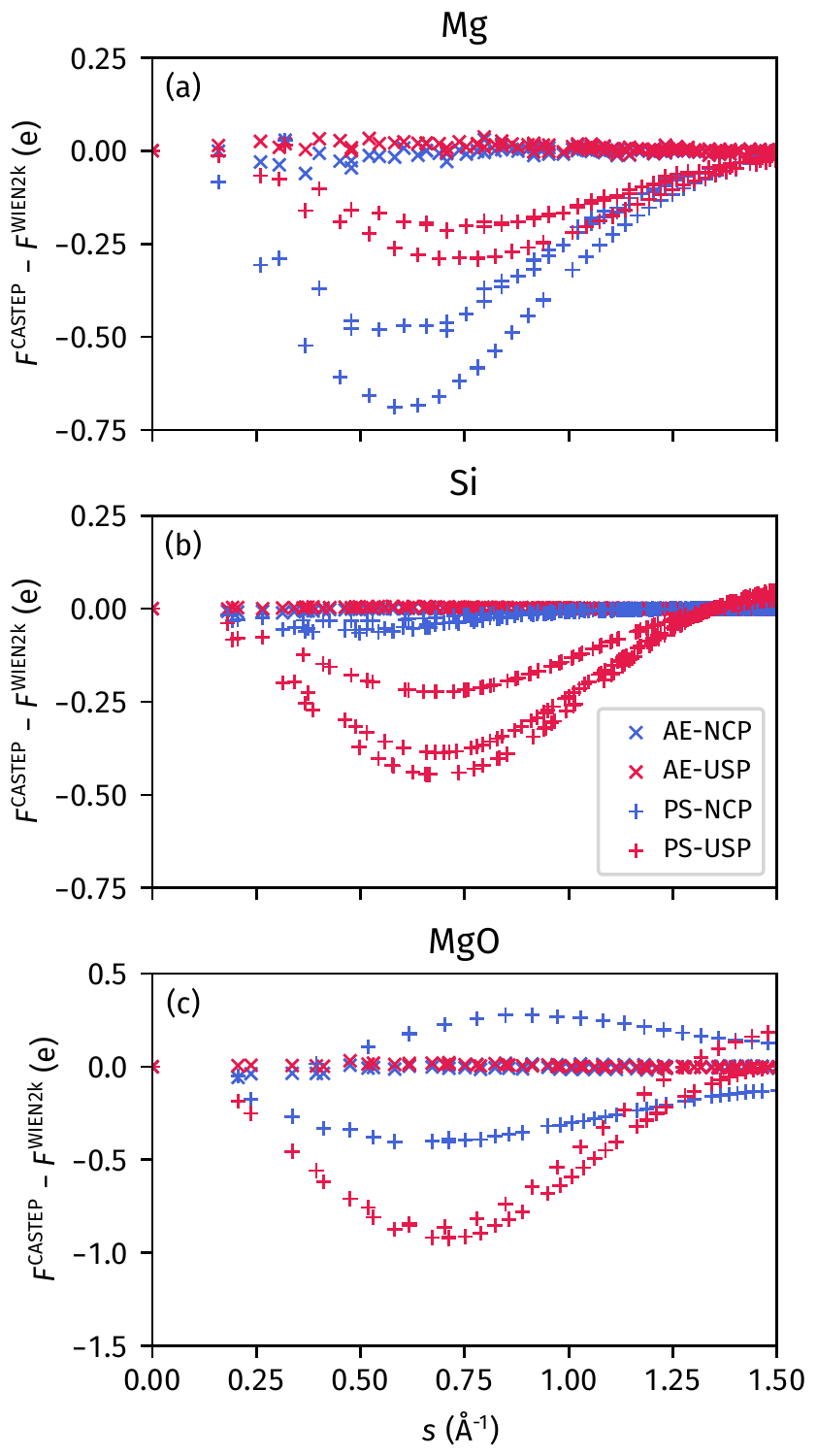}
	\caption{\label{fig:aedftcomparison} The difference between APW+lo WIEN2k SF calculations and those obtained from CASTEP as a function of scattering vector $s$ for (a) Si, (b) Mg and (c) MgO. In CASTEP, PAW DFT calculations were performed employing both USPs and NCPs, with each either using an all-electron (AE) or pseudized (PS) augmentation charge. The PBE functional was used.}
\end{figure} 

In Fig.~\ref{fig:aedftcomparison}, the difference between AE DFT WIEN2k and PAW DFT CASTEP calculations, using AE or PS augmentation charges for both USPs and NCPs, were performed for scattering vectors up to 1.5 \AA{}\textsuperscript{-1} for all three materials systems.
These plots were for the PBE DFA as we found that the LDA mirrored the plots (see Fig.\ S2 of the Supplemental Material~\cite{sm}), suggesting that the differences arise due to the particular pseudization schemes for the pseudopotentials and augmentation charge, which is independent of DFA.
For the two monoatomic systems, we see that in the $s$ range of 0.25-1.25 \AA{}\textsuperscript{-1}, there is a systematic underestimation of the SFs for PAW DFT methods that utilize a pseudized (PS) augmentation charge (e.g., PS-NCP and PS-USP).
This underestimation arises because these high Fourier components are removed when the oscillating valence ED is made smooth. 
The magnitude of this underestimation depends on the exact pseudization schemes employed for the pseudopotentials and augmentation charge (for USP) around each atom. 
Beyond 1.1 \AA{}\textsuperscript{-1}, the SF contributions from the PS valence ED and PS augmentation charge become less than 0.01 e, so that the SF behavior becomes a reflection of the core ED used by the pseudopotential.

\begin{table}[]
	\caption{\label{tab:aedftcompare} $R^\textrm{WIEN2k}$ (\%) values for the PAW DFT method employing either the USP and NCP in combination with either AE or PS augmentation charges.}
	\begin{ruledtabular}
	\begin{tabular}{cccccc}
		&     & AE-USP   & PS-USP    & AE-NCP    & PS-NCP    \\ \hline
		\multirow{2}{*}{Si}  & LDA & 0.05       & 0.56       & 0.04       & 1.03       \\
		& PBE & 0.04       & 0.57       & 0.04       & 0.99       \\
		\multirow{2}{*}{Mg}  & LDA & 0.04       & 3.75       & 0.02       & 0.29       \\
		& PBE & 0.03       & 3.66       & 0.02       & 0.28       \\
		\multirow{2}{*}{MgO} & LDA & 0.05       & 3.26       & 0.08       & 2.01       \\
		& PBE & 0.05       & 3.19       & 0.08       & 1.98      
	\end{tabular}
\end{ruledtabular}
\end{table}

Regardless of the pseudopotential, the deviations resulting from a PS augmentation charge is completely removed when the AE augmentation charge is used.
The $R^\textrm{WIEN2k}$ (in Table \ref{tab:aedftcompare}) lowers by at least an order of magnitude for all studied systems from a $R^\textrm{WIEN2k}$ of 0.28-3.75 \% for PS-USP and PS-NCP to a $R^\textrm{WIEN2k}$ of 0.02-0.08 \% for AE-USP and AE-NCP. 
The errors of AE-USP and AE-NCP with respect to the APW+lo WIEN2k SF values are small enough that they are of the same order as the experimental errors that can arise in high quality x-ray diffraction experiments.
Thus, SFs from AE-USP and AE-NCP (but not PS-USP or PS-NCP) can be used to compare against experiments, as we have done in Sec.~\ref{sec:comparisonexpt}.

MgO is the only system studied here which contains more than one type of atomic species. 
For the PS-NCP calculation, there are two observed behaviors in the SFs, with the $(hkl)$ reflections where $h,k,l$ are all odd overestimating the SFs, while those with $h,k,l$ all even underestimating the SFs. 
This contrasting behavior arises because these two sets arise from different types of reflections. 
Using Eq.~\ref{eq:coreformfactor} for an IAM, the structure factors where $h,k,l$ are all odd take the form: $4(f_\textrm{Mg} - f_\textrm{O})$, arising from a scattering difference between its ions ~\cite{hammondBasicsCrystallographyDiffraction2015a} while those that are all even arises from a constructive summation: $4(f_\textrm{Mg} + f_\textrm{O})$. 
Thus, the SFs with odd $h,k,l$ become sensitive to the specific pseudopotentials of the two atoms, and may overestimate if one pseudopotential underestimates its scattering factor more strongly than the other.

\section{\label{sec:comparisonexpt}Comparison against experiment}

Ultimately, the aim of computing AE SFs from DFT is to use it together with experimental SFs, either to augment the (limited) experimental SFs for better ED reconstructions or to validate the approximations in both experiments and DFT. 
In this section, we will showcase the strength of our approach for the latter application of AE SFs by comparing DFT with two different DFAs and pseudopotentials against experiment, revealing key insights into the effects of these approximations on the accuracy of the ED.

\subsection{Computational details}

\begin{figure*}[t]
	\includegraphics{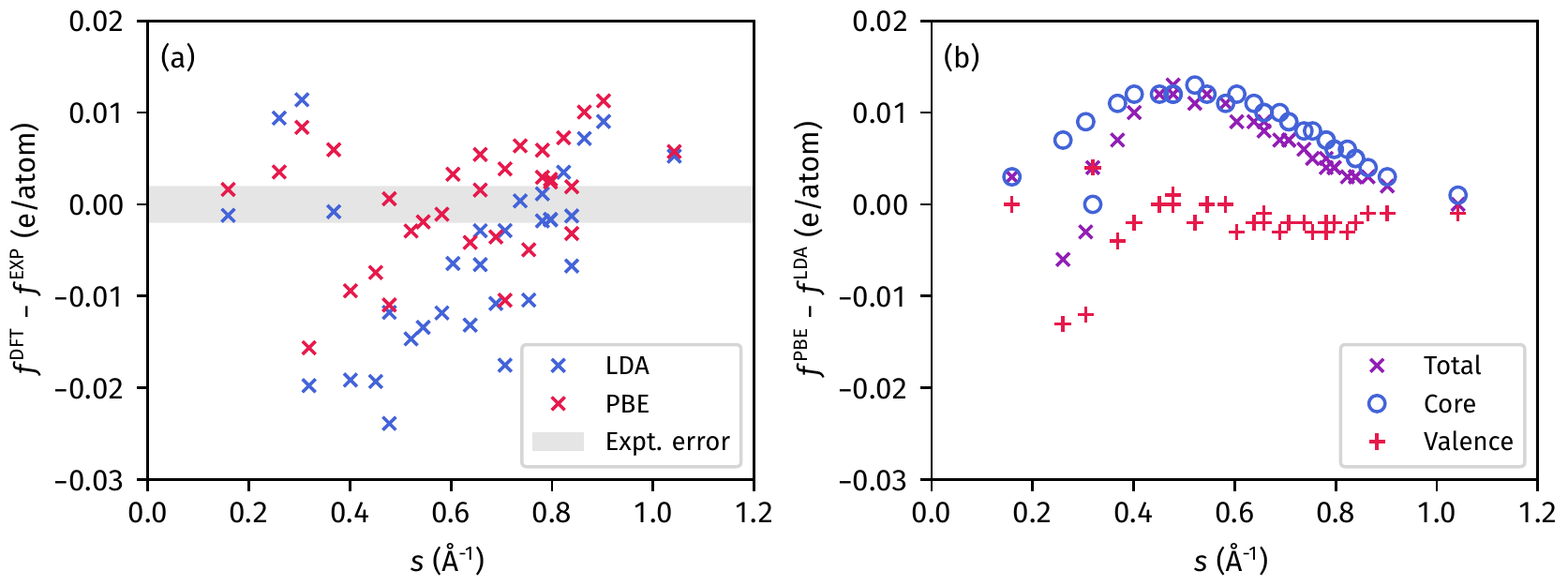}
	\caption{\label{fig:expcomparison} (a) The difference between experimental atomic scattering factors and those obtained from AE-USP CASTEP as a function of scattering vector $s$ for Si for both the LDA and PBE density functional approximations (DFAs). (b) The total difference between the atomic scattering factors of the PBE and LDA DFAs and the separation of this difference into its core and valence (including augmentation) contributions.}
\end{figure*} 

X-ray SFs obtained from diffraction experiments are influenced by the thermal vibrations of the atoms within the crystal~\cite{coppensXrayChargeDensities1997}, tending to reduce the intensity of the diffracted beams. 
Comparison of (static) SFs obtained from DFT requires incorporation of these thermal effects to allow for direct comparison, or alternatively, removal of the thermal effects from experimental numbers.

To a good approximation, the SF is the Fourier transform of the thermally averaged ED $\langle n(\mathbf{r}) \rangle$. 
Within the IAM, thermal effects can be incorporated into the static SF by multiplying each atomic scattering factor $f_j(\mathbf{H})$ in Eq.~\ref{eq:iam3dformfactor} by an isotropic harmonic temperature factor $T_j (\mathbf{H})$:
\begin{equation}
F(\mathbf{H}) = \sum_j \exp(i 2 \pi \mathbf{H} \cdot \mathbf{r}_j)f_j (\mathbf{H}) T_j(\mathbf{H}),
\end{equation}
where
\begin{equation}
T_j(\mathbf{H}) = \exp(-B_j |\mathbf{H}|^2/4).
\end{equation}
The Debye-Waller (DW) factor $B_j$ for atom $j$ can be obtained from either fitting of the static SF to experimental SFs~\cite{lawrenceDebyeWallerFactors1973,valvodaXrayInvestigationAnisotropic1977} or using \textit{ab-initio} phonon dispersion curves~\cite{malicaTemperaturedependentAtomicFactor2019}. 
The derivation of the temperature factor makes two key assumptions: (i) the nuclei vibrate isotropically about their equilibrium positions and (ii) the atomic densities follow the nuclear motion perfectly. 
The second assumption requires that the crystal ED can be divided into a summation of atom-centered density terms.

Within PAW DFT, the second assumption can be applied to the core and augmentation charges, but not the PS valence ED, since it is `delocalized' and cannot be assigned to any one atom. 
For mono-atomic systems, this problem is trivial because the temperature factor can be applied to the static structure factor as a whole, but it cannot be performed for systems with more than one atomic species. 
Likewise, this problem also manifests in the APW+lo AE DFT approach of WIEN2k, where its MTs are localized but the interstitial regions are not. 
Prior studies using the APW+lo codes have overcome this problem by applying the average DW factor of the atomic species -- the average method -- to the interstitial region~\cite{zuoChargeDensityMgO1997b,dudarevCorrelationEffectsGroundstate2000a}; this approximation is sufficient because the interstitial region makes up a small proportion of the total ED. 
The valence electrons in PAW DFT make up a more significant contribution to the total ED compared to the interstitial region in APW+lo AE DFT (see Table S1 of the Supplemental Material~\cite{sm}), such that it may be inaccurate to use an average DW factor. 
Hence, in this study, we will instead use the Hirshfeld partitioning ~\cite{hirshfeldBondedatomFragmentsDescribing1977b} scheme (see Sec.\ S4 of the Supplemental Material~\cite{sm}) to divide the valence electron density into atom-centered densities. 
This approach is found to give a small improvement over the average method in terms of $R^\textrm{EXP}$ across all functionals and pseudopotentials tested for MgO (see Table S2 of the Supplemental Material~\cite{sm}). 
It is expected to give even larger improvements in systems with a large disparity in valence electrons or DW factors where it can account for the relative contributions of the different species.

The SFs with thermal vibrations included through the above scheme was compared to experimental SF for the same three systems as the previous section: Mg~\cite{friisMagnesiumComparisonDensity2003b}, Si~\cite{zuoTheoreticalChargeDensity1997a}, and MgO~\cite{zuoChargeDensityMgO1997b}. 
Experimental details can be found in their respective references.
Thermal effects had already been removed from the structure factors provided by the experimental study on Mg, so we did not apply a Debye-Waller factor to this system.
Compared to theoretical SFs computed from DFT, only a small number of SFs (particularly for Mg and MgO) were available from experiment and only these were compared in the subsequent analysis. 
To be consistent with the prior experimental and theoretical literature on Si, we will compare the `effective' atomic scattering factor of this system. 
Its relation to the structure factor is given by:
\begin{equation}
f(\mathbf{H}) = f(hkl) = \dfrac{F(hkl)}{8 \cos(\frac{\pi}{4}(h+k+l))}.
\end{equation}
This equation was derived for the spherical IAM, which predicts the $(222)$ reflection to be forbidden. 
However, both DFT and experiment predict noticeable intensity in this reflection due to the asphericity of the ED, so it was included as well, with its \textit{atomic} scattering factor defined as $f(222) = F(222)/8$ to give the relative contribution from each atom of the conventional unit cell~\cite{zuoTheoreticalChargeDensity1997a}.

\subsection{Results and discussion}

\begin{table}[t]
	\caption{\label{tab:expcompare} $R^\textrm{EXP}$ (\%) values for the APW+lo AE DFT method and PAW DFT method employing either the USP and NCP in combination with either AE or PS augmentation charges.}
	\begin{ruledtabular}
		\begin{tabular}{ccccccc}
			&                          & APW+lo      & AE-USP    & PS-USP    & AE-NCP     & PS-NCP    \\ \hline
			\multirow{2}{*}{Si}  & LDA & 0.24        & 0.20      & 0.71      & 0.25       & 1.52       \\
			& PBE                      & 0.13        & 0.13      & 0.57      & 0.14       & 1.33       \\
			\multirow{2}{*}{Mg}  & LDA & 0.48        & 0.45      & 2.17      & 0.53       & 0.86       \\
			& PBE                      & 0.36        & 0.34      & 2.03      & 0.42       & 0.74       \\
			\multirow{2}{*}{MgO} & LDA & 0.34        & 0.32      & 1.43      & 0.32       & 0.58       \\
			& PBE                      & 0.32        & 0.30      & 1.26      & 0.35       & 0.53      
		\end{tabular}
	\end{ruledtabular}
\end{table}

Table \ref{tab:expcompare} evaluates $R^\textrm{EXP}$ for Si, Mg and MgO from APW+lo WIEN2k and PAW CASTEP calculations.
We have compared PAW results incorporating either AE or PS augmentation charges. 
For both USPs and NCPs, the use of PS augmentation charges leads to an $R^\textrm{EXP}$ which can be three to five times larger than the AE DFT results.
This large $R^\textrm{EXP}$ is particularly caused by a systematic underestimation of high Fourier components (see Fig.\ S3 of the Supplemental Material~\cite{sm}).
When the AE augmentation charge is used, the resulting $R^\textrm{EXP}$ are all within 0.06 \% of the AE DFT results, with the USPs generally performing better than their NCP counterparts. 
In fact, the USPs even appear to be better than AE DFT by up to 0.04 \% in $R^\textrm{EXP}$ across the three systems and two DFAs. 
However, this improvement is not statistically significant because each experimentally determined SF has an associated error in its value, which we found to propagate to an error of around $\pm 0.05 \%$ in the $R^\textrm{EXP}$ for Si.

In Fig.~\ref{fig:expcomparison}(a), the difference in atomic scattering factors for Si between experiment and the AE-USP method from CASTEP is plotted for LDA and PBE.
We have focused on Si specifically because there are many available experimental SFs which have been consolidated to high precision from multiple studies. 
In general, both LDA and PBE underestimates the SFs in the range of $s=$ 0.4-0.8 \AA{}\textsuperscript{-1}. 
The improved $R^\textrm{EXP}$ for PBE arises within the 0.4-0.6 \AA{}\textsuperscript{-1} region, where its underestimation of experimental SFs is less, agreeing with the observations of Zuo \textit{et al}~\cite{zuoTheoreticalChargeDensity1997a}.

Compared to APW+lo DFT, where a large portion of the valence electrons are treated in the MT spheres, our approach makes it simple (see Eq.~\ref{eq:totalsf}) to obtain the core and valence (PS valence and AE augmentation charge) contributions to the SF; the difference in these two contributions between PBE and LDA are plotted for Si in Fig.~\ref{fig:expcomparison}(b). 
Below 0.4 \AA{}\textsuperscript{-1}, the valence electrons play a significant contribution to the differences in the total scattering factors, but this effect becomes small beyond 0.4 \AA{}\textsuperscript{-1}, which arises because the valence electrons contribute less than 2\% to the total scattering factor at that point (see Fig.\ S4 of the Supplemental Material~\cite{sm}). 
Thus, the marked improvement in the 0.4-0.6 \AA{}\textsuperscript{-1} region of PBE over LDA arises predominantly from a better description of the core electrons. 
This core ED makes up the \textit{frozen-}core in PAW DFT, which, in turn, determines the corresponding pseudopotentials. 
As it is difficult to construct pseudopotentials for more sophisticated DFAs such as hybrid functionals~\cite{yangHybridFunctionalPseudopotentials2018}, the GGA pseudopotential is often used instead.
This uncontrolled approximation will make it problematic to compare the ED or SFs of these more advanced DFAs since they share the same (frozen-)core ED as the DFA used to generate the pseudopotential, so any analysis will not reflect improvements in the core ED. 

\section{\label{sec:conclusions} Conclusion}

In conclusion, we have proposed an efficient approach of obtaining accurate x-ray structure factors for the Vanderbilt ultrasoft pseudopotential and projector augmented wave methods within DFT.
Compared to prior approaches, involving constructing the total ED on a uniform regular grid, this approach circumvents such a need by evaluating the core and augmentation charges on logarithmic radial support grids, significantly reducing the calculation cost and time; thus, extending the range and size of systems that can be studied. 
This approach was implemented in CASTEP and used to study three systems: Si, Mg and MgO. 
Comparison of the SFs to all-electron DFT has shown that it is capable of achieving all-electron accuracy if an all-electron augmentation charge is used for both norm-conserving and ultrasoft pseudopotentials. 
Further comparisons to experimental SFs have shown that our approach can give a detailed comparison of the ED of different DFAs and where their deficiencies might occur (i.e. in the core or the valence electrons). 

\section*{Data availability}
The data that supports the findings of this study are available within the article and its supplemental material. The input and output files associated with this study and all analysis can be found on GitHub at \href{https://github.com/benshi97/Data_XRD_Structure_Factor}{benshi97/Data\_XRD\_Structure\_Factor} and \href{https://mybinder.org/v2/gh/benshi97/Data_XRD_Structure_Factor/HEAD?labpath=analyse.ipynb
}{Binder}.

\section*{Acknowledgments}
The authors acknowledge support from the UKCP Consortium, funded by the Engineering and Physical Sciences Research Council (EPSRC) under grant number EP/P022561/1.
R.J.N. gratefully acknowledges financial support from the EPSRC, Grant No.
EP/L022907/1.

%

\end{document}